Full paper

# An increase in membrane cholesterol by graphene oxide disrupts calcium homeostasis in primary astrocytes

*Mattia Bramini[†ᵽ°*], Martina Chiacchiaretta[†°¢], Andrea Armirotti[¥], Anna Rocchi[†ᵽ], Deepali D. Kale[#], Cristina Martin Jimenez[‡], Ester Vázquez[‡], Tiziano Bandiera[#], Stefano Ferroni[Δ*], Fabrizia Cesca[†$°] and Fabio Benfenati[†ᵽ°]*

[†]Center for Synaptic Neuroscience and Technology and Graphene Labs, Istituto Italiano di Tecnologia, 16132 Genova, Italy; [ᵽ]IRCCS Ospedale Policlinico San Martino, 16132 Genova, Italy; [¥]Analytical Chemistry Lab and Graphene Labs, Istituto Italiano di Tecnologia, 16163 Genova, Italy; [#]PharmaChemistry Line and Graphene Labs, Istituto Italiano di Tecnologia, 16163 Genova, Italy; [‡]Departamento de Química Orgánica, Instituto Regional de Investigación Científica Aplicada (IRICA), Universidad de Castilla La-Mancha, 13071 Ciudad Real, Spain; [Δ]Department of Pharmacy and Biotechnology, University of Bologna, 40126 Bologna, Italy; [$]Department of Life Sciences, University of Trieste, 34127 Trieste, Italy.

°Equal contribution

¢ present address: Department of Neuroscience, Tufts University School of Medicine, Boston, MA, USA.

*Corresponding authors:
Mattia Bramini, PhD; e-mail: mattia.bramini@iit.it
Stefano Ferroni, PhD; e-mail: stefano.ferroni@unibo.it

Keywords: graphene, astrocytes, cholesterol, calcium, proteomics

**Abstract**

The use of graphene nanomaterials (GNMs) for biomedical applications targeted to the central nervous system is exponentially increasing, although precise information on their effects on brain cells is lacking. In this work, we addressed the molecular changes induced in cortical astrocytes by few-layer graphene (FLG) and graphene oxide (GO) flakes. Our results show that exposure to FLG/GO does not affect cell viability or proliferation. However, proteomic and lipidomic analyses unveiled alterations in several cellular processes, including intracellular $Ca^{2+}$ ($[Ca^{2+}]_i$) homeostasis and cholesterol metabolism, which were particularly intense in cells exposed to GO. Indeed, GO exposure impaired spontaneous and evoked astrocyte $[Ca^{2+}]_i$ signals and induced a marked increase in membrane cholesterol levels. Importantly, cholesterol depletion fully rescued $[Ca^{2+}]_i$ dynamics in GO-treated cells, indicating a causal relationship between these GO-mediated effects. Our results indicate that exposure to GNMs alters intracellular signaling in astrocytes and may impact on astrocyte-neuron interactions.

# 1. Introduction

Because of their unique physical and chemical properties, graphene nanomaterials (GNMs) are currently exploited for a range of applications in tissue engineering, regenerative medicine and drug/gene delivery to the central nervous system (CNS).[1,2] The implementation of strategies that make use of GNMs for neurological applications, however, requires a detailed knowledge of the impact of such materials on the biology of the various CNS cell populations. The impact of GNMs on the morphological and functional properties of primary neurons has been extensively described.[3-6] Conversely, a detailed description of the biological interactions between GNMs and glial cells has not been fully addressed yet.

Amongst the various glial subtypes populating the CNS, astrocytes represent 20-40 % of cells.[7] Astrocytes actively contribute to the activity of neural circuits by controlling the dynamics of the perineuronal *milieu* through their capacity to maintain the extracellular ion and neurotransmitter homeostasis.[8,9] Moreover, astrocytes are involved in the structural remodeling of neural networks by directing synapse formation and pruning, regulating brain metabolism and participating in the neurovascular coupling.[10,11] Many of these functions are mediated by dynamic variations in intracellular calcium levels ($[Ca^{2+}]_i$), which follow the paracrine activation of metabotropic receptors by signaling molecules released from cells of neuronal, glial and vascular origin.[12] Of note, alterations of astroglial $[Ca^{2+}]_i$ dynamics are observed under several central nervous system diseases; however, whether this dysregulation plays a pathogenic role or reflects reparative processes following brain damage is still unclear.[10,13-17] These diverse roles can be fulfilled because of the ability of astrocytes to respond to a large variety of environmental cues through alterations of their molecular and functional properties.[18-20] Whether astroglial plasticity can be influenced by exposure to various nanomaterials remains to be established.

A limited number of studies have described the specific effects induced by G-related materials (GRMs) on astrocytes. Single-walled carbon nanotubes (SWCNTs) administered as dispersions, strongly affected the viability of chicken-derived glial cell cultures. On the contrary, functionalized water-soluble SWCNTs induced stellation of mouse cortical astrocytes and increased GFAP expression, in the absence of sizeable effects on cell viability.[21] PEG-SWCNT films had differential effects on astrocytes, with thinner films inducing stellation, and thicker films promoting de-differentiation with cell enlargement, reduced GFAP expression and increased proliferation.[22,23] Graphene oxide (GO) did not affect astrocyte viability *in vitro*.[5,24] We have recently shown that GO impacts on astrocyte homeostatic functions by affecting their ability to buffer extracellular $K^+$ and glutamate, with repercussions on the excitability of co-cultured neurons.[24] Although, the cellular and molecular mechanisms underlying the graphene-mediated changes in the astroglial phenotype remain to be identified, injection of reduced GO flakes in the rodent hippocampus did not induce reactive astrogliosis, indicating that GNMs are well tolerated also *in vivo*.[25]

In this work, we studied the effects of long-term exposure of primary cortical astrocytes to few-layer graphene (FLG) and GO on their proteomic and lipidomic profiles. The analysis identified profound alterations in the intracellular protein networks that control $Ca^{2+}$ signaling, accompanied by a marked increase in cell cholesterol that was causally related to $Ca^{2+}$ dysregulation. The effects of GNMs on astrocyte signaling and activation may impact on the functional astrocyte-neuron interactions.

## 2. Results

### 2.1 Production and characterization of FLG / GO flakes

The GRMs used in the present study were FLG and GO flakes. FLG was prepared by exfoliation of graphite through interaction with melamine by ball-milling treatment.[26] After exfoliation, melamine was removed to obtain stable dispersions of FLG in $H_2O$. GO, provided by the Grupo Antolin Ingeniería (Burgos, Spain), was obtained by oxidation of carbon fibers (GANF Helical-Ribbon Carbon Nanofibres, GANF®). Initial GO suspensions were washed with H$_2$O to remove the presence of acids and were fully characterized. Transmission electron microscopy (TEM) analysis revealed a similar lateral size distribution of FLG and GO, showing high polydispersity of the solutions ranging from 100 to 1,500 nm (**Figure 1a-c**). In both FLG and GO, Raman spectroscopy displayed the two main features of G-based materials: the so-called D and G peaks corresponding to disordered carbon and to the sp$^2$ tangential mode, respectively.[27] The D band appeared at 1347 cm$^{-1}$ and the G peak arose at ~1,576 cm$^{-1}$. The average spectrum of FLG showed an I(D)/I(G) ratio of about 0.26, confirming a low level of defects located at the edges of the micrometer flakes (**Figure 1d**). In contrast, the GO average spectrum displayed broad D and G bands with an I(D)/I(G) ratio of about 0.88. A 2D band (~2,700 cm$^{-1}$) was observed in the FLG structure, while the GO spectrum instead displayed a bump. Accordingly, elemental analysis gave very different results between the two GNMs, as displayed in **Figure 1e**, with an amount of oxygen around 5% (w/w) in FLG and 49% in GO. Finally, thermo-gravimetric analysis (TGA) was used to quantify the degree of functionalization of FLG and GO (**Figure 1f**). In agreement with Raman and elemental analyses, the weight loss observed in FLG at 600 ºC was only 8.5%, confirming the low quantity of oxygen groups generated by the exfoliation process in comparison to GO, whose TGA displayed a weight loss of 49.2%.

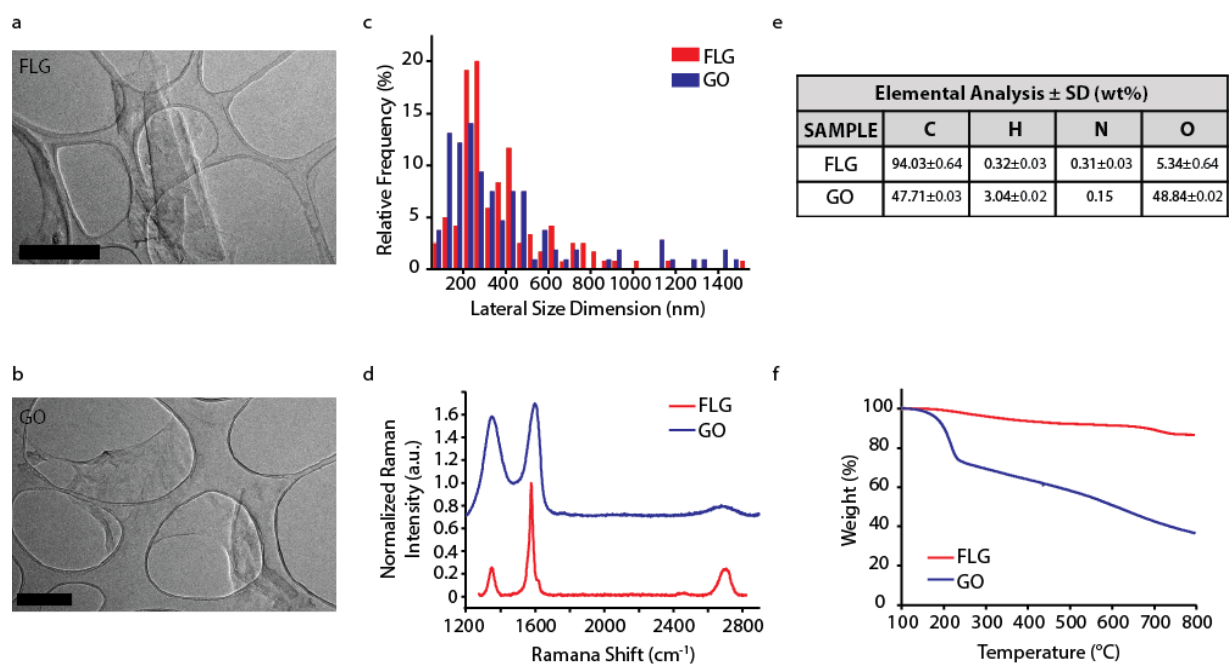

**Figure 1. Characterization of FLG and GO. (a-c)** Representative TEM images of FLG (**a**) and GO (**b**) used throughout the study and their respective lateral size distribution (**c**; n≥107). Scale bars, 500 nm. **(d)** Average Raman spectra (n≥30) of FLG and GO. **(e,f)** Elemental (**e**) and thermo-gravimetric (**f**) analysis of FLG and GO samples.

**2.2 FLG and GO do not affect astrocyte viability and cell cycle and proliferation**

Primary astrocytes were exposed to FLG/GO (1 and 10 μg/ml) or to the corresponding vehicle solution (0.046 ppm melamine/$H_2O$ for FLG or $H_2O$ for GO) for up to 7 days, and cell viability was investigated at various time points by flow cytometry by using propidium iodide (PI) and AlexaFluor488-conjugated Annexin V to detect early and late apoptotic cells, respectively. As a positive control, cells were treated with the strong inducer of apoptosis staurosporine (POS; 1 μM) (**Figure 2a-c**). Cell viability was also assessed after a 2-day exposure of confluent cultures of mature astrocytes to FLG/GO (**Figure 2d**). Exposure to either FLG or GO did not induce cell death at any of the tested concentrations and exposure times. Since no dose-dependent effects were observed on cell viability, all subsequent experiments were performed using the highest GNM concentration (10 μg/ml).

We subsequently evaluated cell division and cell cycle upon exposure to FLG/GO. The cell number in untreated and FLG/GO-treated samples showed comparable changes over time, suggesting a negligible effect of the material on astrocyte division (**Figure 2e**). In accordance with these data, flow-cytometry analysis showed that FLG/GO treatment did not affect the percentage of cells in the various cell cycle phases (**Figure 2f,g**).

We next evaluated whether exposure to GNMs led to astrogliosis (**Figure 2h-k**). To this end, astrocyte cultures were immunostained with antibodies to the intermediate filament glial fibrillary acidic protein (GFAP; **Figure 2h**), a classical marker of reactive gliosis.[28,29] The levels of GFAP were not altered by the exposure to FLG/GO, as shown from both immunostaining (**Figure 2h,i**) and western blot (**Figure 2j,k**) analysis.

These results indicate that exposure to FLG and GO does not affect astrocyte proliferation, survival and reactivity.

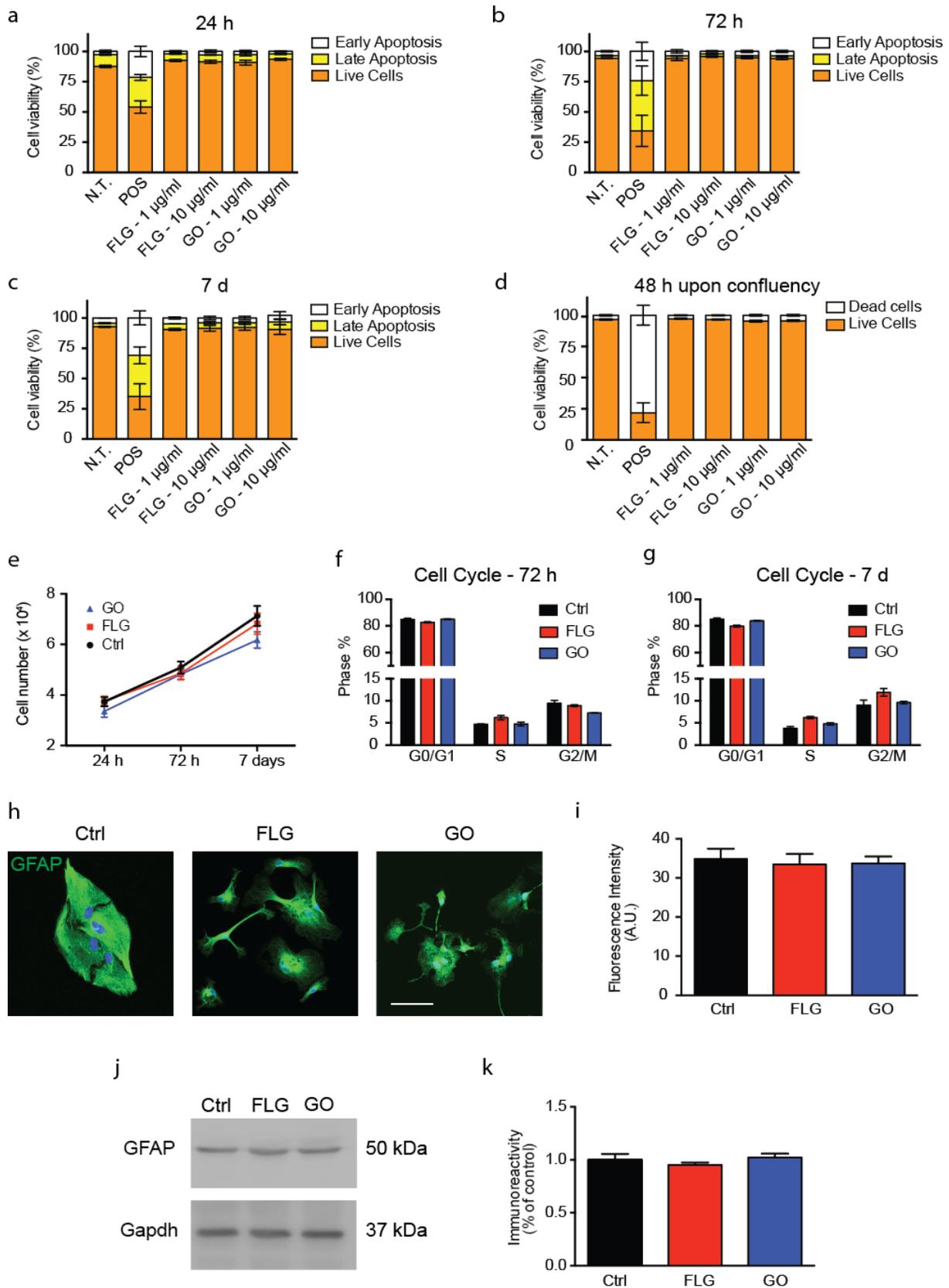

**Figure 2. Exposure to FLG/GO does not alter astrocyte viability, cell cycle and reactivity.**
(**a-c**). Primary rat astrocytes were exposed to either FLG or GO (1 and 10 μg/ml) or to an equivalent volume of the respective vehicle (Ctrl: 0.046 ppm melamine/H$_2$O for FLG, H$_2$O for GO) for 24 h (**a**), 72 h (**b**) and 7 d (**c**). Staurosporine (POS; 1 µM) was used as a positive control for apoptosis. Cell death was evaluated by flow cytometry analysis of Annexin-V (AxV; early apoptotic events) and propidium iodide (PI; late apoptosis) staining. The percentages of AxV- and PI-positive cells with respect to the total number of cells were calculated for each experimental group. (**d**) The same experimental procedure was carried out for primary glial cultures at confluence, exposed for 48 h to either FLG or GO. No changes in cell death were observed under all the experimental conditions. $p>0.05$, Student's *t*-test; 45,000 cells from 3 replicates per experimental condition from n = 3 independent preparations. (**e**) Growth curve of primary astrocytes exposed to FLG and GO for 24 h, 72 h and 7 d. No differences were found compared to vehicle-treated (Ctrl) samples (3 replicates per experimental condition, from n = 3 independent preparations); $p>0.05$, repeated measures ANOVA. (**f,g**) Primary astrocytes were exposed to FLG/GO for 72 h (**f**) and 7 d (**g**). Samples were processed by flow cytometry and analyzed for PI fluorescence intensity. No differences were observed in cell cycle phases across all experimental conditions. $p>0.05$, one-way ANOVA; 45,000 cells from 3 replicates per experimental condition, from n = 3 independent preparations. (**h-k**) GFAP expression upon exposure to FLG/GO. Primary astrocytes were exposed to 10 µg/ml of FLG/GO for 72 h and GFAP expression was quantified by immunofluorescence (**h,i**) and immunoblotting (**j,k**). Representative images (**h**) and quantification of fluorescence intensity (**i**) of astrocytes stained for GFAP (green) and Hoechst (blue) for nuclei visualization. Scale bar, 30 µm. $p>0.05$, one-way ANOVA; 60 cells from n = 3 independent cell preparations. Representative Western blots (**j**) and quantification of the GFAP immunoreactivity (**k**) in astrocytes upon exposure to either FLG or GO. Gapdh was used as loading control. $p>0.05$, one-way ANOVA; n = 3 independent preparations. Data are expressed as means ± sem in all panels.

**2.3 FLG / GO exposure differentially affects the proteomic profile of primary astrocytes**

To get a more comprehensive understanding of the intracellular pathways affected by exposure to GNMs, proteomics (**Figure 3**) and lipidomics (**Figure 4**) screens were carried out on primary astrocytes treated with either FLG or GO for 72 h. We performed untargeted LC-MS/MS-based proteomics by labeling the protein content of astrocytes exposed to GO, FLG or their corresponding vehicles with TMT isobaric tags.[30] Our analysis revealed 255 and 80 proteins significantly altered following GO and FLG treatment, respectively ($p \leq 0.05$). The corresponding Venn diagram analysis showed only 21 proteins shared by the two groups (**Figure 3a**). Gene ontology analysis confirmed the upregulation of key cellular processes. Pathways related to cytoskeletal organization and lipid binding were altered in both FLG and GO-treated samples, although the altered proteins within each category were often different between the two groups (**Figure 3b,c**), in agreement with the Venn diagram. Interestingly, calcium binding and glutamate receptor-binding proteins were specifically altered in GO-treated cells (**Figure 3c**). Changes in the expression of selected Ca$^{2+}$ binding proteins were also validated by RT-PCR (**Figure S1**).

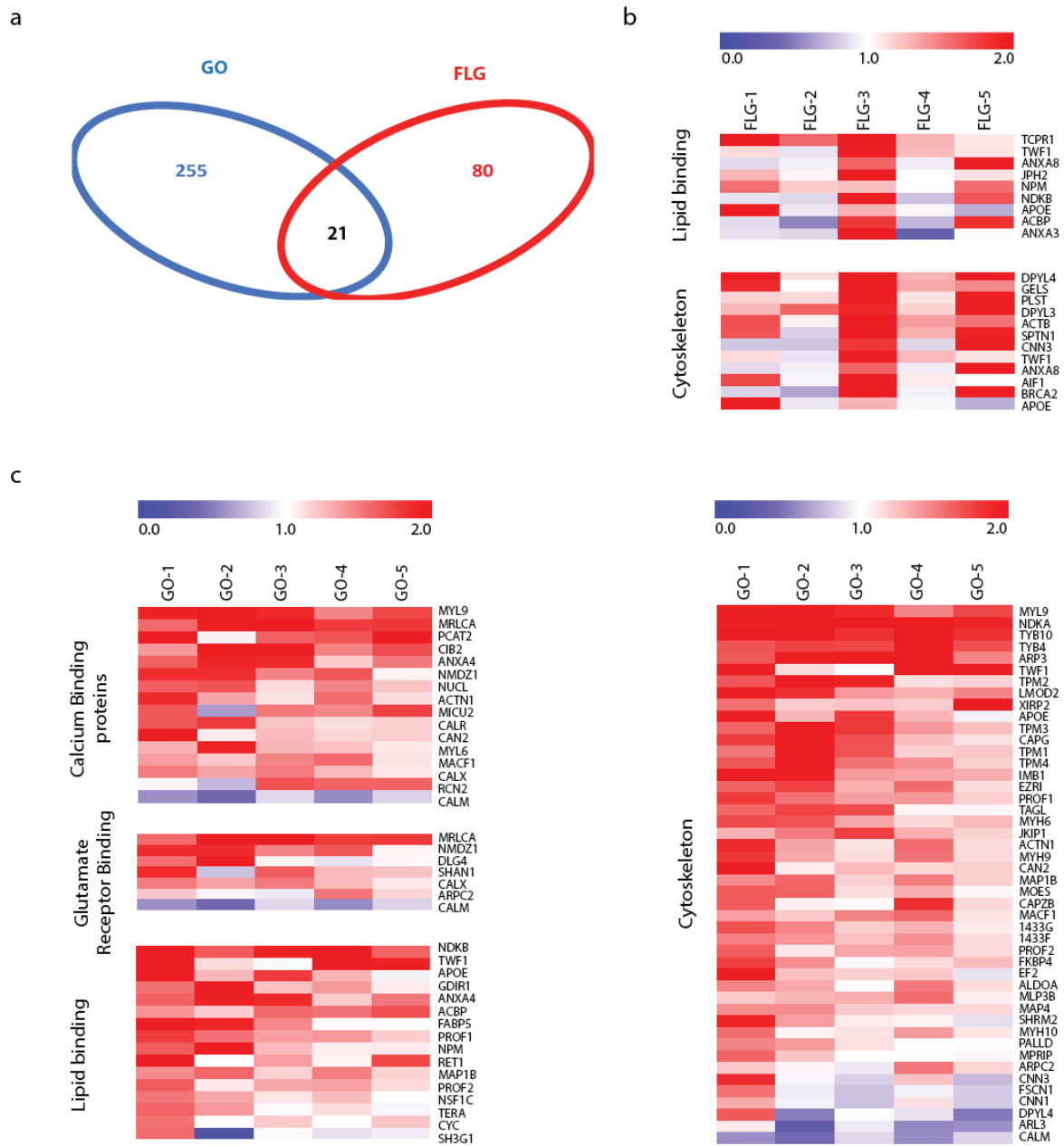

**Figure 3. Proteomic profile of primary astrocytes exposed to FLG/GO.** (**a**) Venn diagram showing the number of unique proteins undergoing significant expression changes upon FLG/GO exposure. Only 21 proteins are in common between the two groups. (**b,c**) Clustered heat-maps of pathway enrichment analysis in primary astrocytes exposed to FLG (**b**) or GO (**c**). n = 5 replicates from 2 animals. P values were calculated using the Benjamini-corrected modified Fisher's exact test and are reported in the **Supplementary Excel Files**.

**2.4 FLG / GO exposure differentially affects the lipidomic profile of primary astrocytes**

Untargeted lipidomics revealed substantial changes in the lipid composition of astrocytes treated with FLG/GO (**Figure 4a,b**). Relevant changes in 46 lipids were found, belonging to several classes, such as phosphatidylethanolamines (PE), phosphatidylinositols (PI), phosphatidylserines (PS), phosphatidylcholines (PC), diacylglycols (DG), triacylglycerols (TG), hexosyl-ceramides (Hex-Cer) and sterols. Amongst these lipid classes, phospholipid, cholesterol and triglyceride levels were significantly altered following exposure of astrocytes to either FLG or GO (**Figure 4c,d; Supplementary Excel Files**). PCs were upregulated by GO and downregulated by FLG, while an opposite modulation was observed for TGs (**Figure 4c**). Interestingly, both cholesteryl esters (such as CE 18:1) and free cholesterol, that are produced by astrocytes in the brain,[31] were markedly upregulated by GO (**Figure 4d**).

Cholesterol plays a crucial role in astrocyte physiology, as it rapidly re-distributes in the cell membrane in response to a variety of stimuli,[32-34] and is linked to lipid raft assembly and $Ca^{2+}$ signaling.[35] To get more insights into the mechanisms underlying the variations of cholesterol levels induced by GO, we stained primary astrocytes with the cholesterol-specific fluorescent marker filipin. In agreement with the lipidomics results, GO-exposed astrocytes displayed strongly increased cholesterol levels with respect to vehicle-treated cells, while FLG induced only a small increase in filipin staining (**Figure 4e**).
Highly concentrated cholesterol in the plasma membrane is a feature of lipid rafts, membrane microdomains characterized by clusters of lipids and a variety of signaling and scaffolding proteins, including ion channels.[36] We evaluated whether FLG/GO exposure affected raft distribution and organization by live labeling GNM-treated and control astrocytes with the fluorescently labeled cholera toxin B-subunit (Vybrant) that recognizes the GM1 ganglioside, a specific marker of membrane rafts (**Figure 4f,g**). Exposure to GO selectively induced an increase in total Vybrant fluorescence (**Figure 4g**, *top left*), average size of the lipid rafts (**Figure 4g**, *top right*) and raft membrane coverage (**Figure 4g**, *bottom left*). Moreover, the size distribution of GO-induced lipid rafts was shifted toward higher values (>10 μm$^2$), indicating that GO-treated cells displayed a higher number of lipid rafts of larger size (**Figure 4g**, *bottom right*).

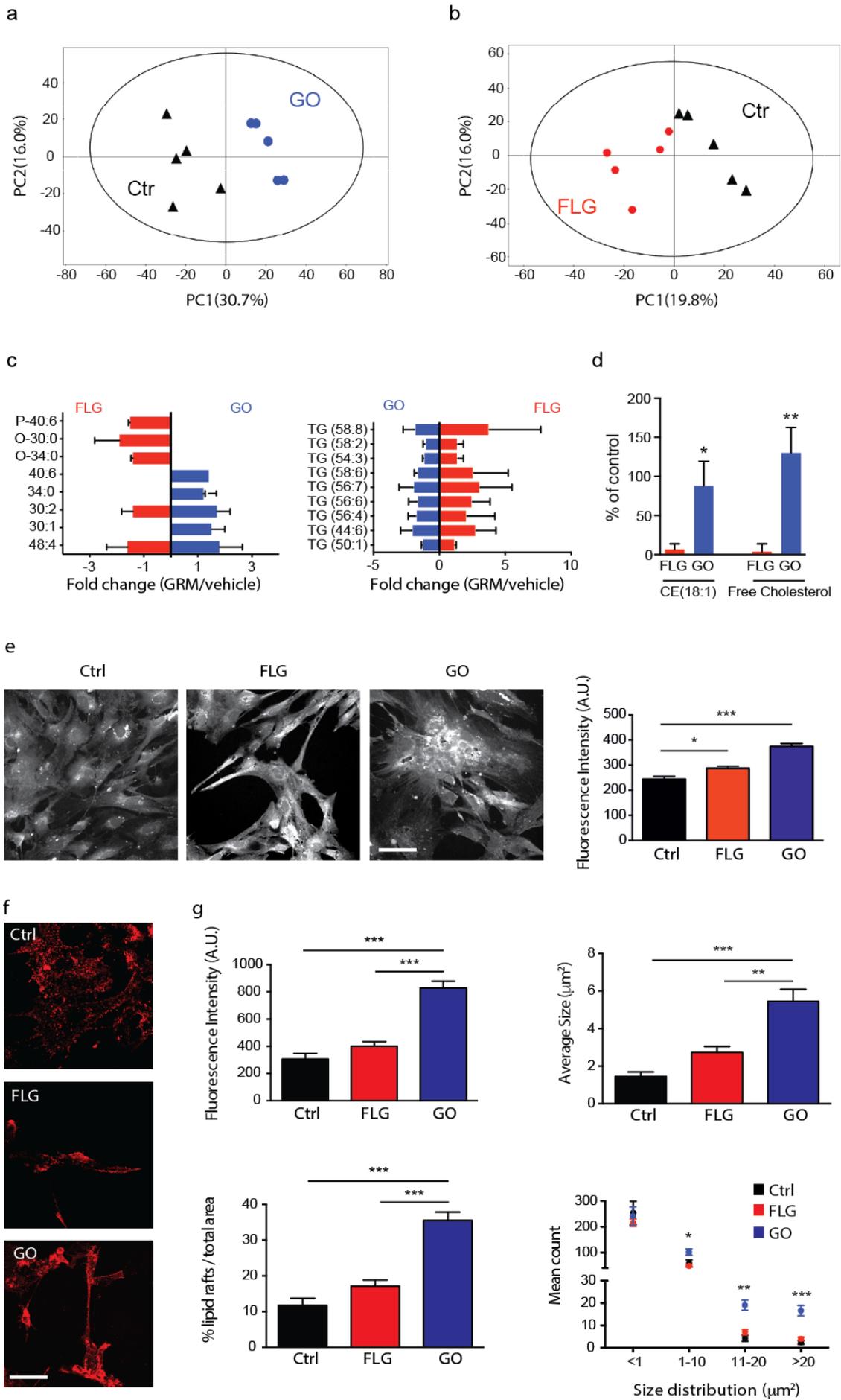

**Figure 4. Lipidomic profiles of primary astrocytes exposed to FLG/GO. (a,b)** Score plots from principal component analysis of untargeted lipidomics data. Dots representing individual astrocyte samples clusterize into two distinct groups (treated/control) for both GO (a) and FLG (b) exposure. (**c**) Changes induced in phosphatidylcholine (*left*) and triacylglycerol (*right*) content following exposure to GO or FLG. (**d**) Changes in oleoyl-cholesterol ester and free cholesterol content in FLG/GO treated astrocytes expressed in percentage of the respective vehicle-treated sample. * p<0.05; ** p<0.01; Student's *t*-test, n=5 replicates from 2 animals. (**e**) Representative images (*left*) and quantification of fluorescence intensity (*right*) of filipin-stained astrocytes after treatment with either vehicle or FLG/GO for 72 h. Scale bar, 20 μm. 30 cells from n = 3 independent cell preparations. (**f,g**) Representative images (**f**) and quantification of the fluorescence intensity of Vybrant-positive rafts (**g**) after treatment with either vehicle or FLG/GO for 72 h. In GO-treated cells, lipid rafts displayed increased mean fluorescence intensity, larger size and larger membrane coverage. n=20 cells, from 2 independent preparations. In (e-g): * p<0.05; ** p<0.01; *** p<0.001, one-way ANOVA/Bonferroni's multiple comparison tests. Data are expressed as means ± sem in all panels.

## 2.5 Exposure to GO inhibits intracellular $Ca^{2+}$ signaling

As described above, treatment with GO specifically perturbed the cellular pathways involved in $[Ca^{2+}]_i$ homeostasis. To gain more insights into the functional consequences of these changes, we exposed primary astrocytes to FLG/GO for 72 h and analyzed $[Ca^{2+}]_i$ dynamics by live imaging of single-cells loaded with the $Ca^{2+}$-sensitive dye FURA2-AM (**Figure 5**). FLG- and GO-treated cultures displayed a reduction in the percentage of cells showing spontaneous $[Ca^{2+}]_i$ transients compared to vehicle-treated cultures, which was accompanied by a reduction in the frequency of spontaneous $[Ca^{2+}]_i$ oscillations and significantly lower basal $[Ca^{2+}]_i$ levels (**Figure 5a**), which reached statistical significance only in GO-treated astrocytes. GO-treated cells also showed significantly impaired ATP-evoked $[Ca^{2+}]_i$ elevations (**Figure 5b**).

To unravel the mechanisms underlying the observed effects, we asked whether GO alters the release of $Ca^{2+}$ from intracellular stores and/or the influx of $Ca^{2+}$ through the plasma membrane.[37] To address this issue, we exposed GO-treated astrocytes to thapsigargin in $Ca^{2+}$-free extracellular saline to deplete intracellular $Ca^{2+}$ stores (**Figure 5c**).[38] Interestingly, GO-treated astrocytes showed a significantly decreased response to thapsigargin, indicating an impairment of the intracellular $Ca^{2+}$ storage/$Ca^{2+}$ release mechanisms. When the extracellular $Ca^{2+}$ concentration was returned to 2 mM to promote $Ca^{2+}$ entry through membrane store-operated $Ca^{2+}$ channels (SOC), GO-treated astrocytes displayed a significantly lower increase in $[Ca^{2+}]_i$ than that evoked in vehicle-treated astrocytes (**Figure 5c**). These results corroborate those of the proteomic screening, showing that exposure of primary astrocytes to GO causes a complex dysregulation of $[Ca^{2+}]_i$ signaling that affects the various mechanisms involved in $[Ca^{2+}]_i$ homeostasis.

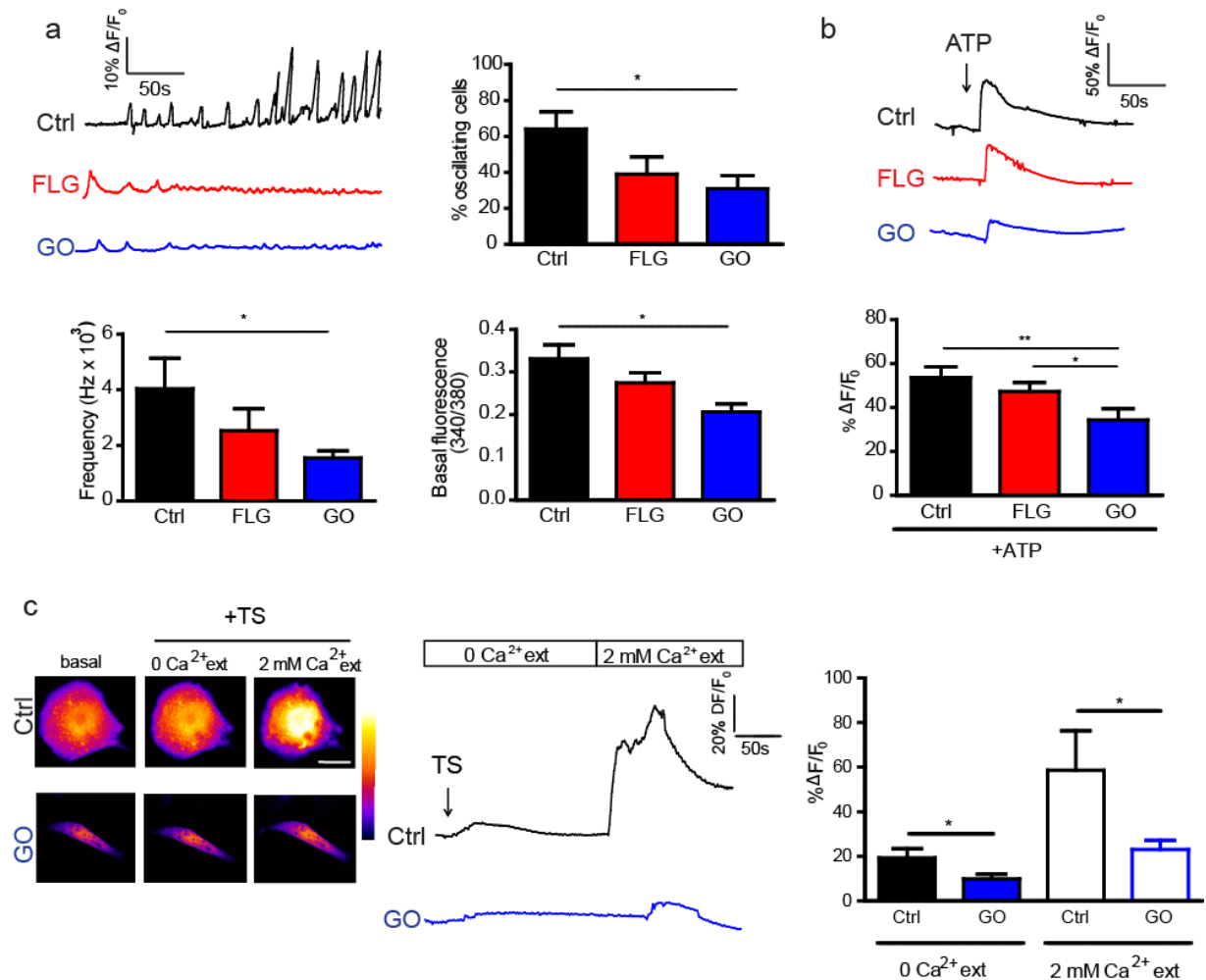

**Figure 5. Alteration of $[Ca^{2+}]_i$ in primary astrocytes treated with FLG/GO.** (a) Representative spontaneous $[Ca^{2+}]_i$ oscillations recorded in astrocytes treated with FLG/GO for 72 h *(top left)*; percentage of spontaneously active cells *(top right)*, frequency of $[Ca^{2+}]_i$ oscillations under basal conditions *(bottom left)* and basal $[Ca^{2+}]_i$ levels *(bottom right)* are shown. *p<0.05, one-way ANOVA/Bonferroni's multiple comparison tests; n = 10 coverslips per experimental condition from 4 independent cell preparations. (b) *Top*: Representative $[Ca^{2+}]_i$ responses evoked by ATP (10 μM) in astrocytes treated for 72 h with FLG/GO and quantification of ΔF/F₀ values *(bottom)* in response to ATP. *p<0.05, **p<0.01 one-way ANOVA/Bonferroni's multiple comparison tests; n = 10 coverslips per experimental

condition from 3 independent preparations (Ctrl, 21; FLG, 25; GO, 21 cells). (**c**) *Left*: Representative images of FURA2-AM-loaded astrocytes exposed to vehicle (Ctrl) or GO under basal conditions and upon addition of thapsigargin (TS; 1 μM), in the absence or presence of 2 mM extracellular $Ca^{2+}$, as indicated. Scale bar, 10 μm. Representative traces of $[Ca^{2+}]_i$ transient peaks (*middle*) and changes in $\Delta F/F_0$ values (*right*) triggered by TS in $Ca^{2+}$-free external medium (0 $Ca^{2+}$ ext) and after replacement of the external solution with a $Ca^{2+}$-containing saline (2 mM $Ca^{2+}$ ext). *$p<0.05$, Mann Whitney *U*-test (n = 10 cells per experimental condition from 3 independent preparations). Data are expressed as means ± sem in all panels.

**2.6 Cholesterol depletion rescues $[Ca^{2+}]_i$ dynamics in GO-treated astrocytes**

A number of studies have suggested that cholesterol plays a regulatory role in $[Ca^{2+}]_i$ homeostasis in various cellular contexts.[39-41] Since cholesterol levels markedly increased in GO-treated cells, we next asked whether the alterations of $[Ca^{2+}]_i$ dynamics induced by GO could be causally linked to the increase in cholesterol content. To address this question, we incubated vehicle- and GO-treated astrocytes with the cholesterol-depleting agent methyl-β-cyclodextrin (MβCD, 7.5 mM) for 30 min and analyzed $[Ca^{2+}]_i$ dynamics (**Figure 6**). Analysis of filipin staining showed that the treatment with MβCD indeed decreased endogenous cholesterol from control astrocytes. Remarkably, in astrocytes exposed to GO, the administration of MβCD decreased filipin staining to levels similar to those of vehicle-treated cells (**Figure 6a**). $[Ca^{2+}]_i$ dynamics upon thapsigargin treatment were not significantly affected by MβCD exposure in vehicle-treated astrocytes (**Figure 6b**, *left* and **c**). By contrast, in GO-treated astrocytes the MβCD incubation rescued $[Ca^{2+}]_i$ signals, which returned to levels comparable to those of vehicle-treated cells (**Figure 6b**, *right* and **c**). Altogether, these data demonstrate the existence of a functional correlation between the increase in cholesterol and the alteration of $[Ca^{2+}]_i$ dynamics induced by GO.

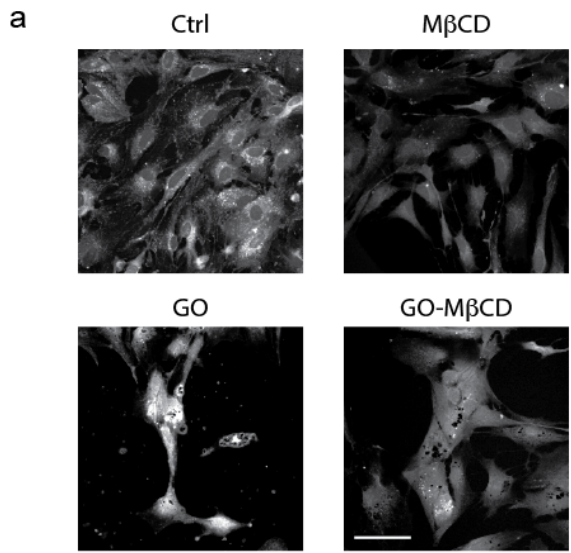
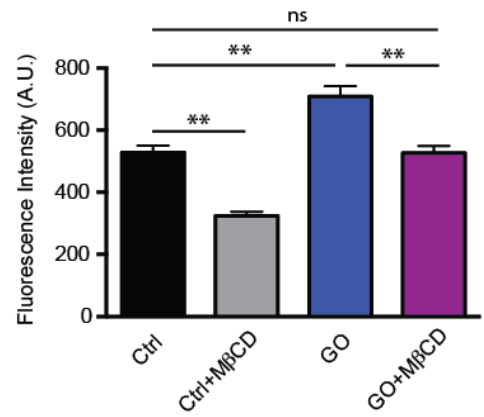
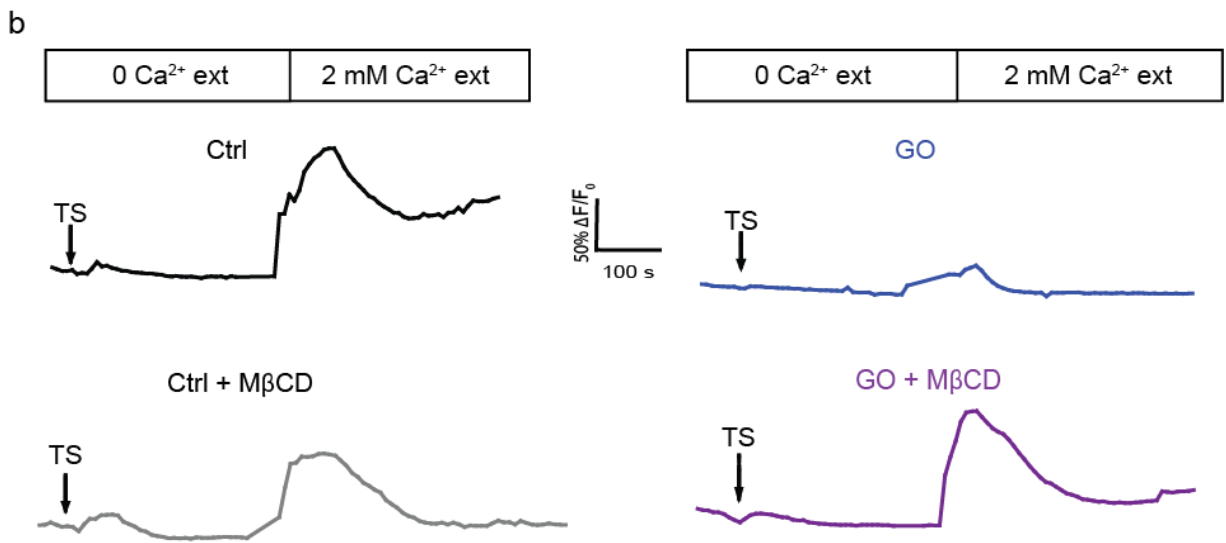
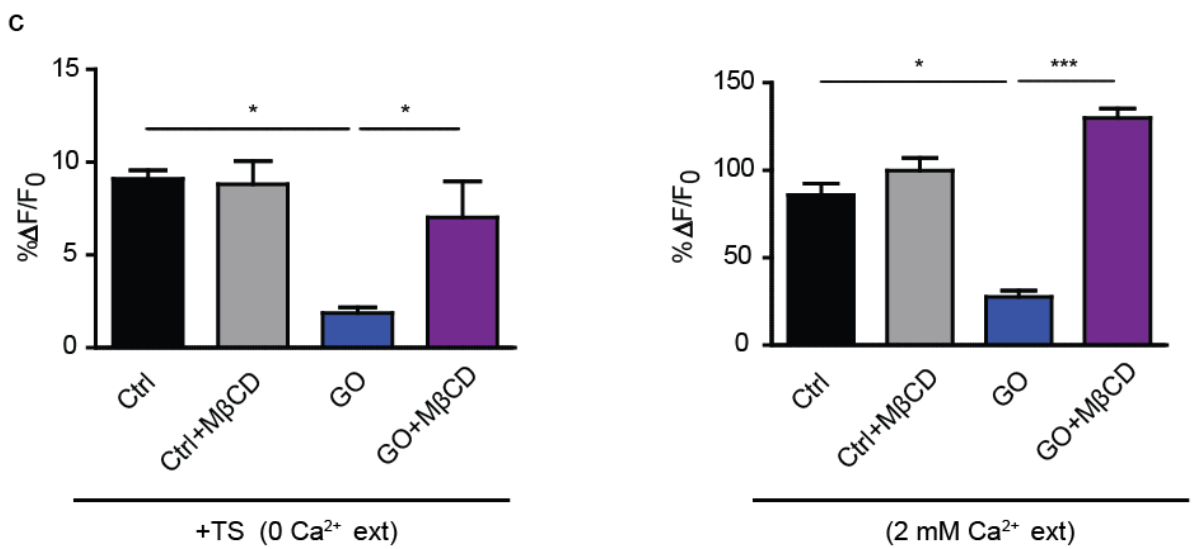

**Figure 6**. Cholesterol depletion rescues GO-induced alterations of $[Ca^{2+}]_i$ dynamics. (**a**) Primary astrocytes that had been exposed to either vehicle or GO for 72 h were incubated in the presence or absence of MβCD for 30 min. Representative images (*left*) and quantification of filipin staining (*right*) used to visualize the effects of the treatments on cholesterol levels. Scale bar, 50 μm. **p<0.01, two-way ANOVA/Bonferroni's multiple comparison tests; n = 8-10 fields per experimental condition from 2 independent preparations. (**b,c**) Representative traces of $[Ca^{2+}]_i$ signals (**b**) and quantification of $\Delta F/F_0$ values (**c**) in primary astrocytes that had been exposed to vehicle (Ctrl, *left*) or GO (*right*) and subsequently incubated in the absence or presence of MβCD for 30 min. Astrocytes were challenged with TS (1 μM) in the absence of extracellular $Ca^{2+}$ (0 $Ca^{2+}$ ext), followed by replacement of the external solution with a $Ca^{2+}$-containing saline (2 mM $Ca^{2+}$ ext). *p<0.05, ***p<0.001; two-way ANOVA/Bonferroni's multiple comparison tests (n ≥ 12 cells per experimental condition from 2 independent preparations). Data are expressed as means ± sem in all panels.

## 3. Conclusion and Discussion

It is well known that astrocytes *in vivo* respond to pathological insults through complex molecular and functional changes that modify their homeostatic properties and alter the structural and functional integrity of the neuronal tissue.[42] These dynamic changes define astrocyte reactivity and can be detrimental or beneficial according to the pathological settings. The upregulation of GFAP expression has been considered a marker of astrocyte reactivity and a landmark of various pathological conditions such as CNS insults, ischemia and several neurodegenerative diseases.[43] Although cultured astrocytes avidly uptake GNMs through the endo-lysosomal pathway,[24] here we show that exposure to GNMs does not affect astrocyte viability and does not cause any significant alteration in cell cycle progression. In addition, the expression of GFAP is unaltered, indicating that, at least *in vitro*, exposure of cortical astrocytes to GNMs does not trigger reactive astrogliosis.

The physico-chemical properties of GNMs are dictated by their size, shape and chemistry, which in turn determine the biological effects they cause upon interaction with living organisms.[44] Our study shows that exposure to GNMs profoundly affects the proteome of cultured astrocytes. While over 300 proteins showed appreciable changes upon GO treatment, only 100 were altered by FLG, and remarkably, only 21 proteins were common to the two materials. This supports the notion that the distinct features of FGL and GO flakes strongly influence their interaction with cultured cells, triggering the activation of distinct, almost non-overlapping pathways. The common processes affected by both FLG and GO include cytoskeleton remodeling and lipid binding proteins. The observed alteration of cytoskeletal proteins is in line with the marked shape change displayed by GNM-treated astrocytes, which acquire a highly elongated morphology reminiscent of activated cells.[24] The impact of GNMs on lipid binding proteins is of interest, as it correlates with the wide array of lipid species that are up- or downregulated in GNM-exposed cells, particularly cholesterol.

The lipidomic screen revealed that phospholipid, cholesterol and triglyceride metabolism is significantly altered following exposure of astrocytes to GO and FLG. PCs, the most abundant structural phospholipids playing a crucial role in the modulation of membrane fluidity,[45] are up-regulated by GO and down-regulated by FLG, while TGs, the main storage form of lipid precursors,[45] show an opposite modulation. Interestingly, cholesteryl esters and free cholesterol, fundamental lipids produced by astrocytes in the brain,[31,46] are markedly upregulated by GO. Brain cholesterol is mainly synthetized *in situ*, since dietary supply is fenced off by the blood-brain barrier. In the adult brain, most of cholesterol is produced by astrocytes in support to the metabolic needs of neurons.[31,46] Thus, the observed increase in cholesterol, in addition to cell-autonomous effects on astrocytes, may impact on the metabolic and electrical activity of co-cultured neurons. Interestingly, and in line with our findings, increased membrane cholesterol levels have been recently reported in primary neurons and cell lines grown onto to graphene substrates. Remarkably, cholesterol was the principal mediator of the cellular and physiological alterations induced by graphene on the overlying cell cultures.[47]

Amongst the pathways specifically altered by GO, a prominent role is played by $Ca^{2+}$-binding proteins. Indeed, analysis of single cell $[Ca^{2+}]_i$ dynamics shows that GO, but not FLG, causes significant alterations of both spontaneous and evoked $[Ca^{2+}]_i$ signals. Moreover, both $Ca^{2+}$ release from intracellular stores and $Ca^{2+}$ entry through the plasma membrane are down-regulated by GO exposure. It is well known that $[Ca^{2+}]_i$ signaling is instrumental for astrocyte homeostatic functions and astrocyte-neuron communication.[48] Notably, alterations of astrocyte $[Ca^{2+}]_i$ dynamics have been associated with a number of CNS dysfunctions.[49] A decrease in ATP- and neurotransmitter-evoked $[Ca^{2+}]_i$ signaling has been reported in astrocytes during aging.[50,51] By contrast, pathological elevation of both spontaneous and evoked $[Ca^{2+}]_i$ responses in astrocytes have been described in a mouse model of Alzheimer's disease.[52] In focal ischemia, the spreading of cell death from the ischemic core to adjacent areas is mediated by $[Ca^{2+}]_i$ waves triggered by the massive release of ATP and activation of P2Y purinergic receptors.[15,53] More recent data have shown that enhanced astrocytic $[Ca^{2+}]_i$ responses actively contribute to some forms of epilepsy.[54,55] Based on this evidence, the demonstration that GO depresses $[Ca^{2+}]_i$ dynamics in primary astrocytes corroborates the view that this material could affect CNS functions *in vivo* also through changes in astrocyte-astrocyte and astrocyte-neuron signaling.

The efficacy of $Ca^{2+}$ signaling in astrocytes relies on the efficient replenishment of the ER store through store-operated $Ca^{2+}$ entry (SOCE). This process is mediated by the dynamic interactions between the ER protein STIM and the plasma membrane protein ORAI that predominantly occur at membrane rafts.[56-58] Cholesterol-rich membrane rafts are essential for the efficient propagation of $Ca^{2+}$ transients, as they assemble the required repertoire of $Ca^{2+}$ channels and receptors to sense external stimuli and translate them into the activation of appropriate intracellular signaling cascades.[59] Recently, a cholesterol-binding domain was identified in STIM1. The binding of STIM1 to cholesterol prevents the interaction with ORAI, and consequently inhibits $Ca^{2+}$ influx from the extracellular environment into the cell.[60] In line with this evidence, our results indicate that cholesterol depletion is sufficient to rescue $[Ca^{2+}]_i$ dynamics in GO-treated cells, while the expression levels of the various SOCE components were not directly affected by GNM exposure (not shown). Thus, by increasing the total amount of membrane cholesterol, GO may inhibit the proper functioning of SOCE and consequently impair the generation and propagation of intracellular $[Ca^{2+}]_i$ waves. Collectively, our current and previously published results[6,24] support the view that GO is an exploitable material for biomedical applications in neuromedicine, since it is not harmful to neurons and astrocytes and regulates important astrocyte homeostatic functions. However, since GO affects the balance between excitatory and inhibitory signaling in neuronal networks[6] and modulates $[Ca^{2+}]_i$ dynamics in primary astrocytes through changes in cholesterol levels, its applicability as functional substrate for prosthetic devices remains to be unequivocally determined. In this context, future studies will address whether the GO-induced effects described here are relevant for the proper functionality of the astrocyte syncytium and neuron-astrocyte communication *in vivo*.

## 4. Experimental Section

*Synthesis and characterization of pristine graphene and graphene oxide*

Few-layer graphene (FLG) flakes were prepared by exfoliation of graphite through interaction with melamine by ball-milling treatment.[61] Elemental analysis revealed that FLG contained 0.46% (w/w) melamine, corresponding to 0.046 ppm for a 10 μg/ml FLG suspension. Graphene oxide (GO) was provided by Grupo Antolin Ingeniería (Burgos, Spain) by oxidation of carbon fibers (GANF Helical-Ribbon Carbon Nanofibres, GANF®) and sodium nitrate in sulfuric acid at 0 °C. For a detailed description of FLG and GO synthesis and physical-chemical characterization, please refer to (Leon *et al.*, 2016)[26] and (Bramini *et al.*, 2016)[6].

*Preparation of primary astrocytes*

All experiments were carried out in accordance with the guidelines established by the European Community Council (Directive 2010/63/EU of 22 September 2010) and were approved by the Italian Ministry of Health (Authorization #306/2016-PR of March 24, 2016). Primary cultures were prepared from wild-type Sprague-Dawley rats (Charles River, Calco, Italy). All efforts were made to minimize suffering and to reduce the number of animals used. Rats were sacrificed by $CO_2$ inhalation, and 18-day embryos (E18) were removed immediately by cesarean section. Briefly, enzymatically dissociated astrocytes were plated on poly-D-lysine-coated (0.01 mg/ml) cell culture flasks and maintained in a humidified incubator with 5% $CO_2$ for 2 weeks. At confluence astrocytes were enzymatically detached using trypsin–EDTA and plated on glass coverslips (Thermo-Fischer Scientific, Waltham, MA) at a density of 20,000-40,000 cells/ml, depending on the experiment. Cultures were incubated at 37 °C, 5% $CO_2$, 90% humidity in medium consisting of DMEM (Gibco/Thermo-Fischer Scientific) supplemented to reach the final concentration of 5% glutamine, 5% penicillin/streptomycin and 10% Fetal Bovine Serum (FBS; Gibco/Thermo-Fischer Scientific). For experiments involving G treatments, cultures were incubated at 1 day *in vitro* (DIV) in a medium containing either 1 or 10 μg/ml of either FLG or GO. Controls were subjected to the same medium change with the addition of equivalent volumes of the respective vehicle (0.046 ppm melamine/$H_2O$ for FLG, $H_2O$ for GO). Cultures were used at DIV 2, 4 and 8 (i.e., after 1, 3 and 7 days of FLG/GO incubation, respectively). All chemicals were purchased from Life Technologies/Thermo-Fischer Scientific unless stated otherwise.

*Cell viability assay and cell cycle analysis*

Primary astrocytes were exposed to FLG or GO (1 and 10 μg/ml), or to equivalent volumes of the respective vehicle (0.046 ppm melamine/$H_2O$ for FLG or $H_2O$ for GO) for 24 h, 72 h and 7 d. The medium was collected in flow-tubes and cells were detached by trypsin EDTA 0.25% (Gibco/Thermo-Fischer Scientific) and re-suspended in 500 μl phosphate-buffered saline (PBS) in the same flow-tubes of the medium. Cells were stained with propidium iodide (PI, 1 μM) for late apoptosis quantification, and AlexaFluor-488 conjugated Annexin V (Molecular Probes, 5 μM) for early apoptosis quantification. Cells were incubated at room temperature (RT) for 15 min in the dark. Cells incubated for 8 h with staurosporin (POS; 1 μM; Sigma Aldrich) were used as positive control for apoptotic cell death. Cell death assays were performed by flow-cytometry analysis using a BD Facs Aria II Cell Sorter (Becton Dickinson, NJ, USA).

For cell counting experiments, primary astrocytes were exposed to 10 μg/ml FLG/GO for 24 h, 72 h and 7 d. Cells were then collected, stained with trypan blue solution (Sigma-Aldrich), and counted using a Neubauer chamber (Sigma-Aldrich).

For cell cycle analysis, primary astrocytes were trypsinized and collected in flow-tubes. Cells were fixed in 70% EtOH for 40 min at -20 °C, centrifuged and the pellet was finally resuspended in 500 μl of 0.2 mg/ml RNase A, 0.05% Triton X-100 and 60 μg/ml PI for 1 h at RT. After the incubation, samples were centrifuged and cell pellet re-suspended in 500 μl PBS for flow analysis. When analyzing samples, the PI fluorescent signal was collected in a linear scale using a FACS Aria II Cell Sorter (Becton Dickinson, NJ, USA). Samples were analyzed at a low flow rate under 400 events/s.

*Immunofluorescence staining*

Primary astrocytes were fixed in PBS /4% paraformaldehyde (PFA) for 20 min at RT. Cells were permeabilized with 1% Triton X-100 for 5 min, blocked with 2% fetal bovine serum (FBS) in PBS/0.05% Tween-80 for 30 min at RT and incubated with primary antibodies in the same buffer for 45 min. The primary antibody used was mouse monoclonal anti-glial fibrillary acidic protein (GFAP, #G3893, Sigma-Aldrich). After the incubation with primary antibodies and several PBS washes, neurons were incubated for 45 min with the secondary antibodies in blocking buffer solution. Fluorescently conjugated secondary antibodies were from Molecular Probes (Thermo-Fisher Scientific; Alexa Fluor488 #A11029). Samples were mounted in ProLong Gold antifade reagent with DAPI (#P36935, Thermo-Fisher Scientific) on 1.5 mm-thick coverslips. Image acquisitions were performed using a confocal microscope (SP8, Leica Microsystems GmbH, Wetzlar, Germany) at 63x (1.4 NA) magnification. Z-stacks were acquired every 300 nm; 10 fields/sample. Offline analysis was performed using the ImageJ software and the JACoP plugin for co-localization studies. For each set of experiments, all images were acquired using identical exposure settings. For cholesterol investigation, Filipin (F-9765, Sigma-Aldrich) staining was carried out. Briefly, cells were fixed as described above and blocked with 1.5 mg/ml of glycine in PBS for 10 min at RT to quench the PFA. Cells were incubated for 2 h at RT with filipin (0.05 mg/ml in 10% FBS-PBS solution). Confocal images were acquired at 63x (1.4 NA) magnification, using 405 nm excitation wavelength and filter-set 430-nm long pass filter.

*Lipidomic and Proteomic Studies*

*Materials.* Solvents and chemicals were purchased from Sigma-Aldrich (Milan, Italy). For proteomics, TMT sixplex kits and C18 spin columns were purchased from Thermo-Fisher Scientific (Waltham, MA, USA). LC-MS instruments and columns, Markerlynx and PLGS software were purchased from Waters Inc. (Milford, MS, USA). The Xcms software (https://xcmsonline.scripps.edu) was used for lipidome data analysis. The Mascot software was purchased from Matrix Science Ltd (London, UK). Lipid identification was carried out using the publicly available database METLIN (https://metlin.scripps.edu/) and LIPIDMAPS (http://www.lipidmaps.org/). Gene ontology and pathway analysis was performed using DAVID Bioinformatics Resources (v 6.8). We used DAVID default population (*Rattus norvegicus*) background in enrichment calculation.[62,63] Heat maps were generated using the MeV TM4 software V 4.9 (http://www.tm4.org). The UNIPROT database (http://www.uniprot.org/) was used to perform manual annotation of biological functions of proteins.

*Sample preparation.* For all the "omics" studies, primary astrocytes were plated in 60 mm Petri dishes pre-coated with 0.01 mg/ml PDL at a density of $1 \times 10^6$ cell/dish, then incubated for 72 h with either 10 μg/ml of FLG or GO and the respective control solutions (see above). Five biological replicates of primary astrocyte cultures from two independent preparations (individual rats) were prepared and used for both proteomics and lipidomics analysis.

*Untargeted Lipidomics*

*Sample preparation.* At the end of the incubation (72 h) cells were scraped, washed with PBS and immediately transferred into 1.5 ml Eppendorf tubes. Cell lipid content was then extracted using a Bligh-Dyer protocol.[64] In brief, 2 ml of 1:2 chloroform/methanol mixture (*v/v*) was added to the vials and vortexed for 30 s. Chloroform (0.5 ml) and water (0.5 ml) were then sequentially added and thoroughly mixed after each addition. The samples were then centrifuged for 15 min at 3,500 *g* at RT. At the end of the process, the organic (lower) phase (~1.5 ml) was transferred to glass vials. The aqueous phases were re-extracted to increase the overall recovery. The organic phases from both extractions were pooled in a glass vial and dried under a nitrogen stream. The extracted lipids were re-dissolved in 0.1 ml of a 9:1 methanol: chloroform solution and analyzed by liquid chromatography coupled to high-resolution mass spectrometry.

*Data acquisition.* Untargeted lipidomics of astrocyte lipid extracts was performed on a UPLC Acquity system coupled to a Synapt G2 QToF high-resolution mass spectrometer. Lipids were separated on a CSH C18 column (1.7 micron particle size, 2.1 x 100 mm). Mobile phase A consisted of acetonitrile: water (60:40) with 10 mM ammonium formate and mobile phase B consisted of acetonitrile: isopropyl alcohol (10:90) with 10 mM ammonium formate. The following gradient program was applied: 15% B for 1 min after injection, then increase to 60% B in 9 min, then to 75% B in 8 min and then to 100% B in further 2.5 min. An isocratic 100% B step was then maintained for 2.5 min and the column was subsequently reconditioned to 15% B for 2 min. Total run time was 25 min with the following conditions: flow rate, 0.4 ml/min; column temperature, 55 °C; injection volume, 6 μl. The instrument was operated in positive ESI mode. MS source parameters were as follows: capillary and cone voltages were set at 2.8 kV and 30 V respectively. Source and desolvation temperature were set at 100 °C and 450 °C respectively. Desolvation gas and cone gas ($N_2$) flows were 800 and 50 L/h, respectively. Mass spectra were recorded in MSe mode with MS/MS fragmentation performed in the trap region on the instrument. Low-energy scans were acquired at a fixed 4 eV collision energy, and high-energy scans were acquired using a collision energy ramp from 20 to 40 eV. Spectra were recorded at a mass resolution of 20,000 in the range of m/z 50 to m/z 1200. The scan rate was set to 0.3 spectra per second. A leucine-enkephalin solution (2 ng/ml) was continuously infused in the ESI source (4 μl/min) and acquired every 30 s for real-time mass axis recalibration. All the samples were run in random order. QC samples, consisting of a pool of all the samples from the four experimental groups, were acquired every five samples and were used to assess system suitability, performance, and reproducibility.

*Data analysis.* Multivariate data analysis of LC-MS features was performed using the Waters Markerlynx 4.1 software with the EZinfo 2.0 (Umetrics, Umea, Sweden) software package. Principal Component Analysis (PCA)[65] was performed. The peak areas of the observed features were normalized by the total ion current and Pareto-scaled prior to PCA. Scores plots were used to visualize differences between challenging GNMs and their corresponding vehicles for each biological replicate. XCMSonline[66] was used for univariate lipidomic data analysis. All LC-MS raw data files were converted to the .netCDF format using the waters DataBridge software. The files were then uploaded to XCMSOnline and processed as a pair group experiment using the default UPLC-High Res POS parameters. Briefly, the centWave method was used for feature detection (m/z tolerance = 15 ppm, min peak width = 2 s, max peak width = 25 s). The obiwarp method was used for retention time correction (profStep = 0.5) and chromatograms were aligned using the following parameters: mzwid 0.01, minfrac 0.5, bw = 2. A data matrix reporting the observed accurate mass and retention times ($t_R$) for all the features was generated and converted into an excel file. Univariate statistical analysis was then performed using a two-tailed unpaired Student's *t*-test between GNMs and the corresponding vehicle groups using Microsoft Excel. Following this procedure, only the features present in both rats were retained. Only the most significant features, showing $p < 0.05$ and an absolute fold change of at least 20% in the comparison were used for further annotation.

*Lipid identification.* Lipid identification was performed by interrogating the web-based algorithms METLIN[67] and LIPIDMAPS[68] using the accurate mass measured for each feature. The following adduct species were searched: $[M+H]^+$, $[M+NH_4]^+$, $[M+H-H_2O]^+$, $[M+Na]^+$, and $[M+K]^+$. A maximum of 5 ppm tolerance in mass accuracy was allowed. Putative IDs were finally verified using class-specific retention time and tandem mass analysis.[69]

*Untargeted Proteomics*

*Sample preparation.* At the end of the 72 h incubation, cells from the four groups (incubated with GO, FLG, and respective controls) were scraped and washed with PBS to remove serum, then lysed by MilliQ water addition. Following the BCA protein assay, 50 μg of proteins were collected from all the samples and transferred to Eppendorf tubes. Samples were then isotopically labeled using the TMT sixplex kit,[30] following the protocol suggested by the vendor. In short: samples were reduced with dithiothreitol, alkylated with iodoacetic acid and then digested overnight with trypsin (1:50 in weight with the protein). At the end of the digestion, peptides were labeled separately using TMT sixplex tags then pooled together. The following labeling strategy was used: reporters 126,127,128 for technical replicates of the challenging GNMs, reporters 129,130,131 for the technical replicates of the corresponding vehicle. Peptides resulting from the overnight digestion were fractionated offline at high pH on a reversed phase C18 spin column (0.1% trimethylamine) using increasing concentrations of acetonitrile. A total of 8 fractions were collected for each experiment. The content from each fraction was dried under vacuum then re-dissolved in 80 μl of 3% ACN with 0.1% formic acid for LC-MS/MS analysis.

*LC-MS/MS analysis.* The tryptic peptide mixture was analyzed using a Synapt G2 QToF instrument equipped with a nanoACQUITY liquid chromatographer and a nanoSpray ion source. Peptides were separated on a BEH nanobore column (75 micron ID x 25 cm length) using a linear gradient of ACN in water from 3 to 55% in 2 h, followed by a washout step at 90% ACN (10 min) and a reconditioning step to 3% for 20 min. Flow rate was set to 300 nl/min. Spray voltage was set to 1.6 kV, cone voltage was set to 28 V, spray gas was set to 0.3 l/min. Survey spectra were acquired over the 50-1600 m/z scan range. Multiply (2+, 3+ and 4+) charged ions between 300 and 1400 m/z were selected as precursors for DDA tandem mass analysis and fragmented in the Trap region of the instrument. Collision energy values were automatically selected by the software using dedicated charge-state dependent CE/m/z profiles. Every 60 s a single leucine-enkephalin (2 ng/ml) MS scan was acquired by the LockMass ion source for spectra recalibration.

*Data analysis.* The acquired raw data files were processed using the PLGS software to recalibrate the mass spectra and generate the precursor-fragments peaklist. Protein identification and quantification was performed by interrogating the SwissProt database using the MASCOT Server software.[70] The search parameters were set as follows: quantification: TMT6plex; fixed modifications: carbamidomethyl (C), TMT6plex (N-term), TMT6plex (K); variable modifications: acetyl (K), acetyl (N-term), deamidated (NQ), methyl (DE), oxidation (M), phospho (ST), phospho (Y); peptide tolerance: 30 ppm; fragment tolerance: 0.5 Da; maximum allowed missed cleavages: 2. At least 2 peptides were required for a positive protein identification and quantification. A significance threshold of p>0.05 was set for protein identification in the MASCOT search. Proteins sharing the same set or subset of peptides were considered as a single entry (MASCOT Protein Families). Protein expression ratio was normalized by the average ratio of all the peptides assigned to proteins.

*Protein Extraction and Western Blotting Analysis*

Total protein lysates were obtained from cells lysed in RIPA buffer (10 mM Tris-HCl, 1 mM EDTA, 0.5 mM EGTA, 1% Triton X-100, 0.1% sodium deoxycholate, 0.1% sodium dodecyl sulfate, 140 mM NaCl) containing protease and phosphatase inhibitor cocktails (Roche, Monza, Italy). The soluble fraction was collected and protein concentration was determined using the BCA Protein Assay Kit (Thermo-Fischer Scientific). For western blotting, protein lysates were denatured at 99 °C in 5X sample buffer (62.5 mM Tris-HCL, pH 6.8, 2% SDS, 25% glycerol, 0.05% bromophenol blue, 5% β-mercaptoethanol, deionized water) and separated on SDS-polyacrylamide gels (SDS-PAGE). The following antibodies were used: mouse monoclonal anti-GFAP (#G3893, Sigma-Aldrich) and rabbit monoclonal anti-GAPDH (#2118, Cell Signaling Technology; MA, USA). Signal intensities were quantified using the ChemiDoc MP Imaging System (Biorad, Hercules, CA, USA).

*RNA Preparation and Real-Time PCR Analysis*

Total RNA was extracted with Trizol (Invitrogen, Carlsbad, CA, USA) according to the manufacturer's instructions, purified using RNeasy MinElute Cleanup Kit (Qiagen, Hilden, Germany), and reverse transcribed into cDNA using the SuperScript IV First-Strand Synthesis System (Invitrogen). Gene expression was measured by quantitative real time PCR using the CFX96 Touch Real-Time PCR Detection System (Biorad). Relative gene expression was determined using the 2-ΔΔCT method. Gusb and Hprt1 were used as reference genes. The list of primers used is provided in **Table S1.**

*Calcium Imaging*

Cultures were loaded for 30 min at 37 °C with 1 μg/ml cell-permeable Fura-2 AM (#F1221, ThermoFisher) in the culture medium. Cells were then washed in recording buffer (140 NaCl, 4 KCl, 2 MgCl$_2$, 2 CaCl$_2$, 10 HEPES, 5 glucose, pH 7.4, with NaOH) for 30 min at 37 °C to allow hydrolysis of the esterified groups. Coverslips with cells were mounted on the imaging chamber and loaded with 0.5 ml of recording buffer. Fura-2-loaded cultures were observed with an inverted epifluorescence microscope Leica DMI6000 (Leica Microsystems GmbH, Wetzlar, Germany) using a 63x (1.4 NA) oil-immersion objective, and recordings were performed from visual fields containing 5 ± 2 astrocytes on average. During the analysis, we selected the cells by drawing regions of interest (ROI) around their bodies to reduce any background. Samples were excited at 340 nm and 380 nm by a mercury metal halide lamp (LEICA EL6000, Leica Microsystems). Excitation light was separated from the emitted light using a 395 nm dichroic mirror. Images of fluorescence emission > 510 nm were acquired continuously for a maximum of 2400 s (200 ms individual exposure time) by using an Orca-ER IEEE1394 CCD camera (Hamamatsu Photonics, Hamamatsu City, Japan). The imaging system was controlled by the integrated imaging software package (Leica LAS AF software, Leica Application Suite Advance Fluorescence, version 3.3, Leica Microsystems). In order to monitor spontaneous Ca$^{2+}$ oscillations, astrocytes were recorded for 15 min under baseline conditions. To evoke Ca$^{2+}$ transients, astrocytes were stimulated, after 1 min of baseline acquisition, with 10 μM of ATP (#A2383, Sigma-Aldrich, Milan, Italy). For thapsigargin (#1138, Tocris, Avonmouth, Bristol, UK) experiments, cells were first perfused with a standard Ca$^{2+}$-free extracellular solution (Ca$^{2+}$ was replaced with 5 mM EGTA); after 30 s of baseline acquisition, thapsigargin (1 μM) was added. After [Ca$^{2+}$]$_i$ returned to baseline levels, CaCl$_2$ (2 mM) was added back to the extracellular medium to induce Ca$^{2+}$ entry. For cholesterol depletion experiments, astrocytes were pretreated with methyl-beta-cyclodextrin (MβCD, 7.5 mM; C4555 Sigma-Aldrich) for 30 min before proceeding with Fura2-AM incubation. Fluorescence intensity ratios (R) were calculated for the time course of the experiment according to the equation below:

$$R = \frac{MV\ (\lambda_x 340 - \lambda_e 510) - MV\ background}{MV\ (\lambda_x 380 - \lambda_e 510) - MV\ background}$$

where MV (mean value) is the average pixel intensity for each ROI, and $\lambda_x$-$\lambda_e$ is the excitation (collection) wavelength. For quantitative analysis, the percentage increase in fluorescence ($\Delta F$) was calculated by subtracting the baseline fluorescence ($F_0$) from the total fluorescence (F) and normalizing the difference by the F0 values ($\Delta F/F_0$).

*Statistical Analysis*

Data are expressed as means ± sem throughout. Normal distribution was assessed using D'Agostino-Pearson's normality test. In case of two experimental groups, data were analyzed using the Student's *t*-test or the Mann-Whitney's *U*-test. When more than 2 experimental groups with normally distributed data were involved, one-way or two-ways ANOVA was used, followed by the Bonferroni's multiple comparison test. Significance level was preset to p<0.05. Statistical analysis was carried out using the Prism6 software (GraphPad Software, Inc.).

**Supporting Information**
Supporting Information is available from the Wiley Online Library or from the author.

**Acknowledgements**

The Antolin Group is acknowledged for providing the commercial material. A. Mehilli is gratefully acknowledged for primary cell culture preparations, as well as D. Moruzzo, F. Canu and I. Dallorto for technical and administrative support. The work was supported by the European Union's Horizon 2020 Research and Innovation Programme under Grant Agreement No. 696656 - Graphene Flagship - Core1 and Grant Agreement No. 785219 - Graphene Flagship - Core2.

**Competing financial interests**

The authors declare no competing financial interests.

# Supporting Information

**An increase in membrane cholesterol by graphene oxide disrupts calcium homeostasis in primary astrocytes**

*Mattia Bramini[†ᵇ°\*], Martina Chiacchiaretta[†°¢], Andrea Armirotti[¥], Anna Rocchi[†ᵇ], Deepali D. Kale[#], Cristina Martin Jimenez[‡], Ester Vázquez[‡], Tiziano Bandiera[#], Stefano Ferroni[Δ\*], Fabrizia Cesca[§† °] and Fabio Benfenati[†ᵇ°]*

[†]Center for Synaptic Neuroscience and Tecnology and Graphene Labs, Istituto Italiano di Tecnologia, 16132 Genova, Italy; [ᵇ]IRCCS Ospedale Policlinico San Martino, 16132 Genova, Italy; [¥]Analytical Chemistry Lab and Graphene Labs, Istituto Italiano di Tecnologia, 16163 Genova, Italy; [#]PharmaChemistry Line and Graphene Labs, Istituto Italiano di Tecnologia, 16163 Genova, Italy; [‡]Departamento de Química Orgánica, Instituto Regional de Investigación Científica Aplicada (IRICA), Universidad de Castilla La-Mancha, 13071 Ciudad Real, Spain; [Δ]Department of Pharmacy and Biotechnology, University of Bologna, 40126 Bologna, Italy; [$]Department of Life Sciences, University of Trieste, 34127 Trieste, Italy.

°Equal contribution

¢ present address: Department of Neuroscience, Tufts University School of Medicine, Boston, MA, USA.

*Corresponding authors:
Mattia Bramini, PhD; e-mail: mattia.bramini@iit.it
*Stefano Ferroni, PhD; e-mail: stefano.ferroni@unibo.it*

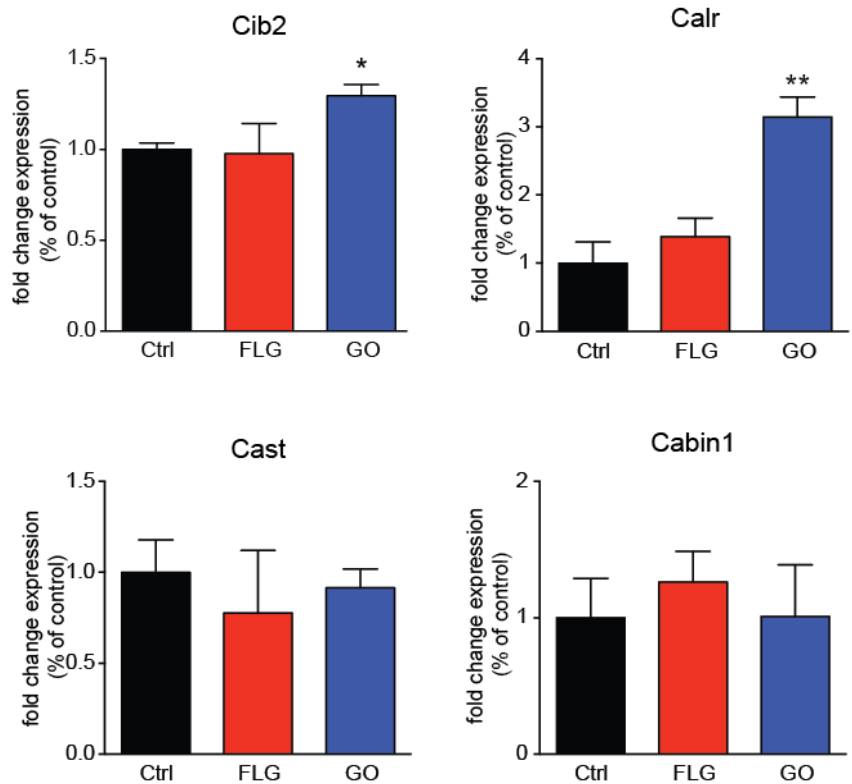

**Figure S1.** mRNA levels of $Ca^{2+}$ and integrin binding family member 2 (*Cib2*), calreticulin (*Calr*), calpastatin (*Cast*) and calcineurin binding protein 1 (*Cabin1*) were quantified by qRT-PCR in primary astrocytes treated with FLG or GO at 10 μg/ml for 72 h. Gusb and Hprt1 were used as reference genes. *$p<0.05$, **$p<0.01$; one-way ANOVA/Bonferroni's tests; n=6. Data are expressed as mean ± sem

**Table S1.** List of primers used for qRT-PCR.

| Gene Symbol | Forward sequence | Reverse sequence |
| --- | --- | --- |
| Cib2 | AGATGAGGTAGTGCTTGT | TCAGATTCGGATGTGGAA |
| Calr | GGTGTTTGTCTTTAATTCTTCA | TTCCTCCATACCTGTTCC |
| Cast | CCTTAGATTTCAAATGCTACC | TCTCTTGTCAGTGCCTTA |
| Cabin1 | GAGGAGGAAGATGATCCA | TCTCAGATTCCATGAATGTG |